\documentstyle[12pt,a4]{article}
\input epsf

\global\arraycolsep=2pt

\begin{document}

\begin{titlepage}

\begin{flushright}
CERN-TH/96-336\\
hep-ph/9612298
\end{flushright}

\vspace{0.5cm}
\begin{center}
\Large\bf
Renormalization of Velocity-Changing\\
Dimension-Five Operators\\
in the Heavy-Quark Effective Theory
\end{center}

\vspace{1.5cm}
\begin{center}
G. Amor\'os$^*$ and M. Neubert\\
{\sl Theory Division, CERN, CH-1211 Geneva 23, Switzerland}
\end{center}

\vspace{1.0cm}
\begin{abstract}
We study the renormalization of operators of the type $\bar
h_{v'}\Gamma\,G^{\mu\nu} h_v$ in the heavy-quark effective theory
(HQET). We construct the combinations of such operators that are
renormalized multiplicatively, and calculate their
velocity-dependent anomalous dimensions at the one-loop order. We
then show that the virial theorem of the HQET is not renormalized,
and that in the limit of equal velocities the anomalous dimension of
the chromo-electric operator vanishes to all orders in perturbation
theory. This implies an exact relation between renormalization
constants, which may help in a future calculation of the two-loop
anomalous dimension of the chromo-magnetic operator.
\end{abstract}

\vspace{1.0cm}
\centerline{(Submitted to Physics Letters B)}

\vspace{2.5cm}
\noindent
CERN-TH/96-336\\
December 1996

\vspace{1.5cm}
\centerline{$^*$\small
On leave from: Departament de F\'\i sica Te\`orica, Universitat de
Val\`encia, Spain}

\end{titlepage}

\section{Introduction}

The heavy-quark effective theory (HQET) is a convenient tool to
explore the physics of hadrons containing a heavy quark
\cite{review}. It provides a systematic expansion around the limit
$m_Q\to\infty$, in which new symmetries of the strong interactions
arise, relating the long-distance properties of many observables to a
small number of hadronic matrix elements. In the HQET, a heavy quark
inside a hadron moving with four-velocity $v$ is described by a
velocity-dependent field $h_v$ subject to the constraint
$\rlap/v\,h_v=h_v$. This field is related to the original heavy-quark
field by a phase redefinition, so that it carries the ``residual
momentum'' $k=p_Q-m_Q v$, which characterizes the interactions of the
heavy quark with gluons. The effective Lagrangian of the HQET is
\cite{EiHi}--\cite{FGL}
\begin{equation}
   {\cal L}_{\rm eff} = \bar h_v\,i v\!\cdot\!D\,h_v
   + \frac{1}{2 m_Q}\,\bar h_v(i D_\perp)^2 h_v
   + \frac{C_{\rm mag}(\mu)}{4 m_Q}\,
   \bar h_v\,\sigma_{\mu\nu} G^{\mu\nu} h_v + O(1/m_Q^2) \,,
\label{Leff}
\end{equation}
where $D_\perp^\mu=D^\mu-(v\cdot D)\,v^\mu$ contains the components
of the gauge-covariant derivative orthogonal to the velocity, and
$G^{\mu\nu}=i[D^\mu,D^\nu]$ is the gluon field strength tensor. The
leading term in the effective Lagrangian, which gives rise to the
Feynman rules of the HQET, is invariant under a global $SU(2N_Q)$
spin--flavour symmetry group. This so-called heavy-quark symmetry
results from the fact that in the limit $m_Q\to\infty$ the properties
of the light constituents inside a heavy hadron become independent of
the spin and flavour of the heavy quark. The symmetry is explicitly
broken by the higher-dimensional operators arising at order $1/m_Q$,
whose origin is most transparent in the rest frame of the heavy
hadron: the first operator corresponds to the kinetic energy
resulting from the motion of the heavy quark inside the hadron (in
the rest frame, $(i D_\perp)^2$ is the operator for $-{\bf k}^2$),
and the second operator describes the magnetic interaction of the
heavy-quark spin with the gluon field. The Wilson coefficient $C_{\rm
mag}(\mu)$ results from short-distance effects and depends
logarithmically on the scale at which the chromo-magnetic operator is
renormalized \cite{EiHi,FGL}. As a consequence of the so-called
reparametrization invariance of the HQET (an invariance under
infinitesimal changes of the velocity), the kinetic operator is not
multiplicatively renormalized \cite{LuMa,Chen}.

One of the most important applications of heavy-quark symmetry
concerns the analysis of semileptonic weak decays of heavy hadrons.
These processes are mediated by flavour-changing currents containing
two heavy-quark fields, which in the HQET are represented by
velocity-changing operators of the type $\bar h_{v'}\Gamma\,h_v$,
where $\Gamma$ represents some Dirac matrix. In the heavy-quark
limit, all weak decay form factors parametrizing the current matrix
elements between two heavy mesons or baryons are proportional to a
universal (Isgur--Wise) form factor $\xi(v\cdot v')$ \cite{Isgu}.
Large logarithms of the heavy-quark masses, which arise due to
quantum corrections, can be summed up to all orders in perturbation
theory by calculating the anomalous dimension of the operator $\bar
h_{v'}\Gamma\,h_v$ in the HQET and solving its renormalization-group
equation (RGE) \cite{Falk,KoRa}.

In the analysis of the symmetry-breaking corrections to the relations
between form factors, higher-dimensional current operators play an
important role. At order $1/m_Q$, there appear local dimension-four
operators of the type $\bar h_{v'}\Gamma\,iD^\mu\,h_v$, whose matrix
elements have been studied in \cite{Luke}. The renormalization of
such operators is completely determined by reparametrization
invariance \cite{RPIMN}. In the present paper, we study the
renormalization of dimension-five current operators containing the
gluon field, i.e.\ operators of the form $\bar
h_{v'}\Gamma\,G^{\mu\nu} h_v$. There are several motivations for this
study: The matrix elements of these operators determine part of the
$1/m_Q^2$ corrections to heavy-hadron weak decay form factors
\cite{FaNe}, which at the present level of accuracy already have to
be included in some applications of the HQET, such as the extraction
of the Cabibbo--Kobayashi--Maskawa matrix element $|V_{cb}|$
\cite{Vcb}. The same operators also play a role in the description of
non-factorizable corrections in non-leptonic two-body decays of $B$
mesons \cite{Blok}. Our primary motivation, however, is the
connection between velocity-changing dimension-five operators and the
velocity-conserving operators appearing at order $1/m_Q$ in the
effective Lagrangian (\ref{Leff}). Clearly, for
$\Gamma=\sigma_{\mu\nu}$ and in the limit of equal velocities, the
operator $\bar h_{v'}\Gamma\,G^{\mu\nu} h_v$ reduces to the
chromo-magnetic operator. More interestingly, however, there is also
a connection with the kinetic operator $\bar h_v(iD_\perp)^2 h_v$. It
is provided by the virial theorem of the HQET, which relates the
kinetic energy of a heavy quark inside a hadron to its interactions
with gluons \cite{virial}. For the ground-state mesons and baryons,
this theorem can be written in the form
\begin{equation}
   \lim_{v'\to v}\,\frac{\langle H(v')|\,\bar h_{v'} v_\mu v'_\nu\,
    i G^{\mu\nu} h_v\,|H(v)\rangle}{(v\cdot v')^2-1} = \frac 13\,
   \langle H(v)|\,\bar h_v(iD_\perp)^2 h_v\,|H(v)\rangle \,.
\label{virial}
\end{equation}
It has been used to study the properties of the kinetic operator under
renormalization, and to estimate its matrix element using QCD sum
rules \cite{MN96}.

Below we study first the general structure of operator mixing of
current operators containing the gluon field and show that it can be
described by a $4\times 4$ matrix of renormalization constants. We
then calculate the corresponding anomalous dimensions at the one-loop
order. Next, we investigate the limit of equal velocities, in which
case the operator basis consists of a chromo-electric and a
chromo-magnetic operator, which are renormalized multiplicatively. We
show that, as a consequence of the virial theorem and
reparametrization invariance, the chromo-electric operator is not
multiplicatively renormalized to all orders in perturbation theory.
This, in turn, implies an exact relation between renormalization
constants, which may help in the calculation of the yet unknown
two-loop anomalous dimension of the chromo-magnetic operator.
Finally, we consider the hadronic matrix elements of
velocity-changing dimension-five operators and determine, for the
case of the ground-state heavy mesons, the scale dependence of the
corresponding hadronic form factors.

\section{Operator mixing and anomalous dimensions}

Our goal is to study the renormalization of the local operator
$O_1^{\mu\nu}=\bar h_{v'}\Gamma\,G^{\mu\nu} h_v$, where $v$ and $v'$
are the heavy-quark velocities, and $\Gamma$ may be an arbitrary
Dirac matrix. Since the Feynman rules of the HQET do not involve
$\gamma$ matrices, the structure of $\Gamma$ will not be altered by
radiative corrections. Under renormalization, the operator
$O_1^{\mu\nu}$ mixes with other operators carrying the same global
quantum numbers. We use the background-field formalism \cite{Abbo}
and work in dimensional regularization, so that it suffices to
consider gauge-invariant operators of the same dimension as
$O_1^{\mu\nu}$. Moreover, we shall not consider operators that vanish
by the equations of motion, since they have vanishing matrix elements
between physical states. To construct the operator basis, we find it
convenient to introduce the two vectors
\begin{equation}
   v_+ = \frac{v+v'}{\sqrt{2(w+1)}} \,,\qquad
   v_- = \frac{v-v'}{\sqrt{2(w-1)}} \,,
\end{equation}
where $w=v\cdot v'$ is the product of the two velocities. This
definition is such that $v_+^2=1$, $v_-^2=-1$, and $v_+\cdot v_-=0$.
Hence, $v_+$ can be regarded as a four-velocity, whereas $v_-$ is a
space-like four vector. In the Breit frame, where the two hadrons
move with opposite velocities, we have $v_+^\mu=(1,{\bf 0})$ and
$v_-^\mu=(0,{\bf n})$, where ${\bf n}$ is a spatial unit vector.
Since the original operator $O_1^{\mu\nu}$ is invariant under
Hermitean conjugation followed by an interchange of the velocities,
the basis operators must contain even powers of $v_-$. Four such
operators can be constructed, and we define
\begin{eqnarray}
   O_1^{\mu\nu} &=& \bar h_{v'}\Gamma\,G^{\mu\nu} h_v \,,
    \nonumber\\
   O_2^{\mu\nu} &=& v_+^{[\mu} v_{+\alpha}\,\bar h_{v'}\Gamma\,
    G^{\alpha\nu]} h_v \,, \nonumber\\
   O_3^{\mu\nu} &=& v_-^{[\mu} v_{-\alpha}\,\bar h_{v'}\Gamma\,
    G^{\alpha\nu]} h_v \,, \nonumber\\
   O_4^{\mu\nu} &=& (v_+^\mu v_-^\nu - v_+^\nu v_-^\mu)\,v_{+\alpha}
    v_{-\beta}\,\bar h_{v'}\Gamma\,G^{\alpha\beta} h_v \,.
\label{basis}
\end{eqnarray}
We use a short-hand notation such that $a^{[\mu} b^{\nu]} = a^\mu
b^\nu - a^\nu b^\mu$. In principle, there are other operators
carrying the same quantum numbers, which contain two derivatives
acting on the heavy-quark fields. However, the Feynman rules of the
HQET ensure that all operators that can mix with the above ones and
do not involve the gluon field strength tensor vanish by the
equations of motion.

We define renormalization constants $Z_{ij}$, which absorb the
ultraviolet (UV) divergences in the matrix elements of the bare
operators, by the relation
\begin{equation}
   O_{i,{\rm bare}}^{\mu\nu} = \sum_{j=1}^4\,Z_{ij} O_j^{\mu\nu} \,.
\end{equation}
From the definition of the basis operators, it follows that the
$4\times 4$ matrix ${\bf\hat Z}=(Z_{ij})$ can be expressed in terms
of the four entries $Z_{1j}\equiv Z_j(w)$, which in general are
functions of the variable $w$. Substituting for $\Gamma$ the
appropriate expressions corresponding to the operators in
(\ref{basis}), we find that
\begin{equation}
   {\bf\hat Z} = \left( \begin{array}{cccc}
   Z_1(w) & Z_2(w) & Z_3(w) & Z_4(w) \\
   0 & Z_1(w) + Z_2(w) & 0 & Z_3(w) + Z_4(w) \\
   0 & 0 & Z_1(w) - Z_3(w) & Z_2(w) - Z_4(w)  \\
   0 & 0 & 0 & Z_1(w) + Z_2(w) - Z_3(w) - Z_4(w)
   \end{array} \right) \,.
\label{Zmat}
\end{equation}
Let us denote by $z_n(w)$ the eigenvalues of this matrix, given by
the diagonal entries. Since the mixing of $O_1^{\mu\nu}$ with
$O_3^{\mu\nu}$ and $O_4^{\mu\nu}$ must vanish in the limit of equal
velocities, it follows that
\begin{equation}
    Z_3(1) = Z_4(1) = 0 \,.
\label{Z34}
\end{equation}
This implies $z_1(1)=z_3(1)$ and $z_2(1)=z_4(1)$. The eigenoperators
${\cal O}_n^{\mu\nu}$, which are renormalized multiplicatively
according to ${\cal O}_{n,{\rm bare}}^{\mu\nu} = z_n(w) {\cal
O}_n^{\mu\nu}$, are given by
\begin{eqnarray}
   {\cal O}_1^{\mu\nu} &=& O_1^{\mu\nu} - O_2^{\mu\nu}
    + O_3^{\mu\nu} - O_4^{\mu\nu} \,, \nonumber\\
   {\cal O}_2^{\mu\nu} &=& O_2^{\mu\nu} + O_4^{\mu\nu}
    \,, \nonumber\\
   {\cal O}_3^{\mu\nu} &=& O_3^{\mu\nu} - O_4^{\mu\nu}
    \,, \nonumber\\
   {\cal O}_4^{\mu\nu} &=& O_4^{\mu\nu} \,.
\label{eigen}
\end{eqnarray}
The anomalous dimensions of these operators, which appear in the RGE
\begin{equation}
   \left( \mu\,\frac{\mbox{d}}{\mbox{d}\mu} + \gamma_n(w) \right)
   {\cal O}_n^{\mu\nu}(\mu) = 0 \,,
\label{RGE}
\end{equation}
are obtained from the relation
\begin{equation}
   \gamma_n(w) = -2\alpha_s\,\frac{\partial}{\partial\alpha_s}\,
   z_n^{(1)}(w) \,,
\label{gamma}
\end{equation}
where $z_n^{(1)}(w)$ denotes the coefficient of the $1/\epsilon$ pole
in $z_n(w)$ calculated in dimensional regularization, i.e.\ in
$d=4-2\epsilon$ space-time dimensions.

\begin{figure}[htb]
\epsfxsize=12cm
\centerline{\epsffile{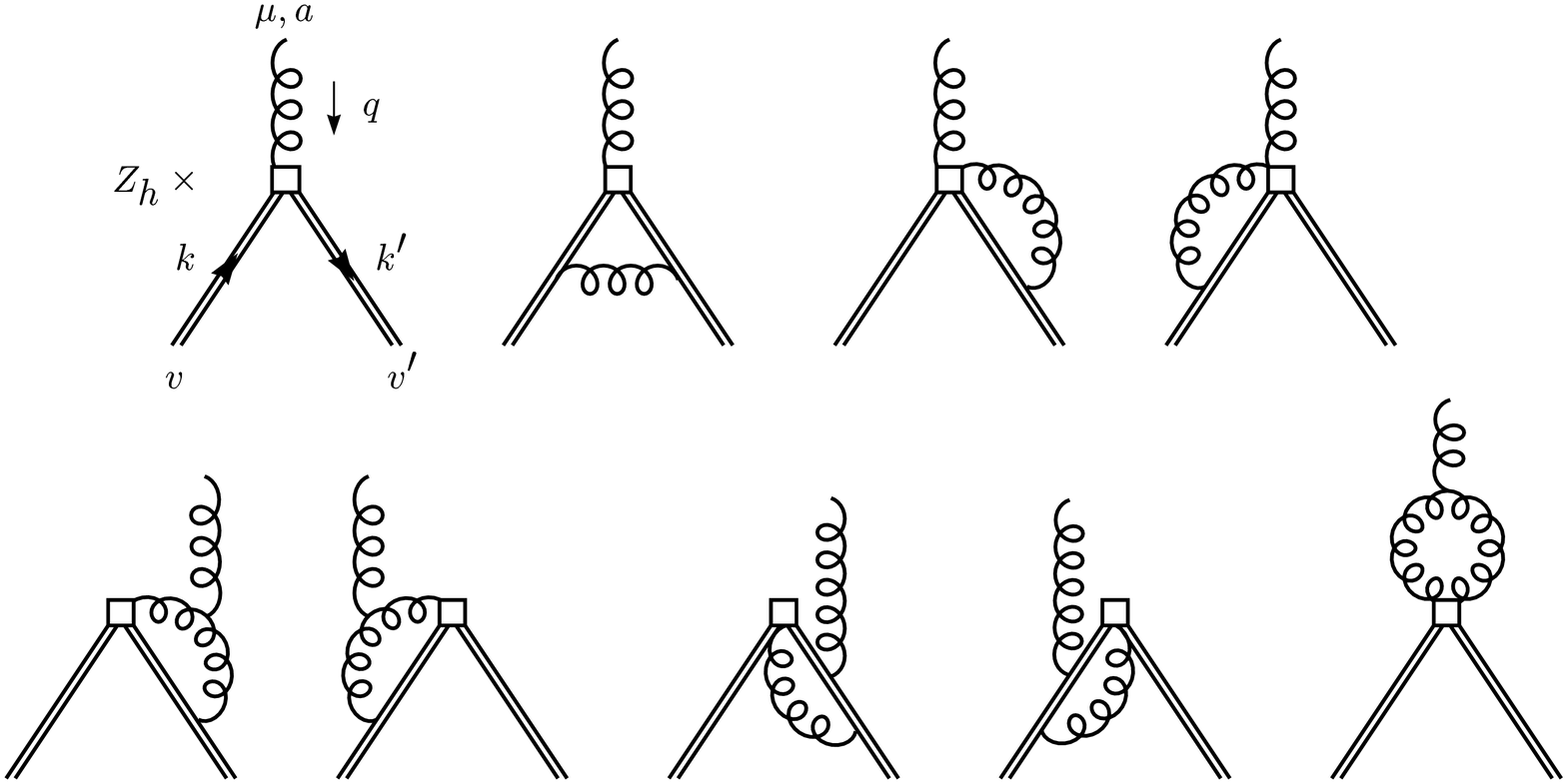}}
\centerline{\parbox{14cm}{\caption{\label{fig:diagrams}
One-loop diagrams contributing to the calculation of the
renormalization factors $Z_n(w)$. The velocity-changing operators are
represented by a square, and heavy-quark propagators are drawn as
double lines.}}}
\end{figure}

At the one-loop order, the renormalization factors $Z_i(w)$ are
determined by the UV divergences of the diagrams shown in
Fig.~\ref{fig:diagrams}. $Z_h$ denotes the wave-function
renormalization constant for the heavy-quark field in the HQET. A
virtue of the background field formalism is that the gluon field is
not renormalized, since $Z_g Z_A^{1/2}=1$ \cite{Abbo}. We have
performed the calculation of these diagrams in an arbitrary covariant
gauge, and with arbitrary momentum assignments. The sum of the UV
divergent contributions is independent of the gauge and of the
external momenta of the heavy quarks. We find
\begin{eqnarray}
   Z_1(w) &=& 1 + \frac{\alpha_s}{4\pi\epsilon}\,\left\{
    (C_A - 2 C_F)\Big[ w\,r(w)-1 \Big] - C_A \right\} \,,
    \nonumber\\
   Z_2(w) &=& \frac{w+1}{2}\,\frac{C_A\alpha_s}{4\pi\epsilon}
    \,, \nonumber\\
   Z_3(w) &=& \frac{w-1}{2}\,\frac{C_A\alpha_s}{4\pi\epsilon}
    \,, \nonumber\\
   Z_4(w) &=& 0 \,,
\end{eqnarray}
where $C_F=\frac 12(N-1/N)$ and $C_A=N$ are the eigenvalues of the
quadratic Casimir operator in the fundamental and the adjoint
representations, and $N$ is the number of colours. The function
$r(w)$ is given by
\begin{equation}
   r(w) = \frac{1}{\sqrt{w^2-1}}\,\ln\left(w+\sqrt{w^2-1}\right)
\end{equation}
and satisfies $r(1)=1$. Note that $Z_3(w)$ vanishes in the limit of
equal velocities, in accordance with (\ref{Z34}). Moreover, in that
limit we find
\begin{equation}
    Z_1(1) + Z_2(1) = 1 \,.
\label{Z12}
\end{equation}
Below we shall argue that this is an exact relation, valid to all
orders in perturbation theory.

Applying now the relation (\ref{gamma}) to the combinations of
renormalization factors appearing in the diagonal entries in
(\ref{Zmat}), we obtain for the one-loop anomalous dimensions
\begin{eqnarray}
   \gamma_1(w) &=& \gamma_4(w) + \frac{C_A\alpha_s}{2\pi} \,,
    \nonumber\\
   \gamma_2(w) &=& \gamma_4(w) - \frac{w-1}{2}\,
    \frac{C_A\alpha_s}{2\pi} \,, \nonumber\\
   \gamma_3(w) &=& \gamma_4(w) + \frac{w+1}{2}\,
    \frac{C_A\alpha_s}{2\pi} \,, \nonumber\\
   \gamma_4(w) &=& -(C_A - 2 C_F)\,\frac{\alpha_s}{2\pi}\,
    \Big[ w\,r(w)-1 \Big] \,.
\end{eqnarray}
In the limit of equal velocities, we find
\begin{eqnarray}
   \gamma_1(1) &=& \gamma_3(1) = \frac{C_A\alpha_s}{2\pi} \,,
    \nonumber\\
   \gamma_2(1) &=& \gamma_4(1) = 0 \,.
\label{zerorec}
\end{eqnarray}
Because of (\ref{Z34}) and (\ref{Z12}), the second relation is valid
to all orders in perturbation theory.

Knowing the one-loop anomalous dimensions allows us to solve the RGE
(\ref{RGE}) for the eigenoperators in the leading logarithmic
approximation. The solution reads
\begin{equation}
   {\cal O}_n(m_Q) = \left( \frac{\alpha_s(m_Q)}{\alpha_s(\mu)}
   \right)^{\gamma_n^0(w)/2\beta_0} {\cal O}_n(\mu) \,,
\label{LO}
\end{equation}
where the coefficients $\gamma_n^0(w)$ are defined by
$\gamma_n(w)=\gamma_n^0(w)(\alpha_s/4\pi)$, and $\beta_0=\frac{11}{3}
N-\frac 23 n_f$ is the first coefficient of the $\beta$ function.

\section{Equal-velocity limit}

The discussion of operator mixing becomes more transparent in the
limit of equal velocities (i.e.\ $v'=v$ and $w=1$), in which $v_+\to
v$ and $v_-\to n$, where $n$ is an external space-like four-vector
satisfying $n^2=-1$ and $n\cdot v=0$. Because of relation
(\ref{Z34}), the mixing of the operators $O_1^{\mu\nu}$ and
$O_2^{\mu\nu}$ decouples from that of $O_3^{\mu\nu}$ and
$O_4^{\mu\nu}$; indeed, the latter two operators are simply
proportional to the first two: $O_3^{\mu\nu}=n^{[\mu} n_\alpha
O_1^{\alpha\nu]}$ and $O_4^{\mu\nu}=n^{[\mu} n_\alpha
O_2^{\alpha\nu]}$. The operators that are multiplicatively
renormalized can be chosen as
\begin{eqnarray}
   O_{\rm mag}^{\mu\nu} &=& O_1^{\mu\nu} - O_2^{\mu\nu}
    = \bar h_v\Gamma\,G_\perp^{\mu\nu} h_v \,, \nonumber\\
   O_{\rm el}^{\mu\nu} &=& O_2^{\mu\nu}
    = v^{[\mu} v_\alpha\,\bar h_v\Gamma\,G^{\alpha\nu]} h_v \,,
\end{eqnarray}
where $G_\perp^{\mu\nu} = (g_\alpha^\mu - v_\alpha v^\mu)
(g_\beta^\nu - v_\beta v^\nu) G^{\alpha\beta}$ contains the
components of the field strength tensor in the subspace orthogonal to
the velocity. In the rest frame, where $v^\mu=(1,{\bf 0})$, the
operators $O_{\rm mag}^{\mu\nu}$ and $O_{\rm el}^{\mu\nu}$ correspond
to purely chromo-magnetic and chromo-electric field configurations,
respectively. The anomalous dimensions of these operators are given
by
\begin{eqnarray}
   \gamma_{\rm mag} &=& \gamma_1(1) = \gamma_3(1)
    = -2\alpha_s\,\frac{\partial}{\partial\alpha_s}\,Z_1^{(1)} \,,
    \nonumber\\
   \gamma_{\rm el} &=& \gamma_2(1) = \gamma_4(1)
    = -2\alpha_s\,\frac{\partial}{\partial\alpha_s}\,
    \Big[ Z_1^{(1)} + Z_2^{(1)} \Big] \,,
\label{gamZ}
\end{eqnarray}
where $Z_n^{(1)}$ is the coefficient of the $1/\epsilon$ pole in
$Z_n(1)$. The anomalous dimension of the chromo-magnetic operator is
known at the one-loop order \cite{EiHi,FGL}, and it is in agreement
with our result for $\gamma_1(1)$ in (\ref{zerorec}). The virial
theorem (\ref{virial}) implies that the chromo-electric operator has
the same anomalous dimension as the kinetic operator $\bar
h_v(iD_\perp)^2 h_v$. However, reparametrization invariance enforces
that to all orders in perturbation theory the kinetic operator is not
multiplicatively renormalized \cite{LuMa}. Therefore, we obtain the
exact relation
\begin{equation}
   \gamma_{\rm el} = \gamma_{\rm kin} = 0 \,,
\label{gamel}
\end{equation}
which is equivalent to (\ref{Z12}). Again, this constraint is
satisfied by our explicit one-loop result for $\gamma_2(1)$ in
(\ref{zerorec}). Using this relation, we obtain
\begin{equation}
    \gamma_{\rm mag} = -2\alpha_s\,
    \frac{\partial}{\partial\alpha_s}\,Z_1^{(1)}
    = 2\alpha_s\,\frac{\partial}{\partial\alpha_s}\,Z_2^{(1)} \,.
\label{nice}
\end{equation}
Recalling the definition of the renormalization factors,
\begin{equation}
   [\bar h_v\Gamma\,G^{\mu\nu} h_v ]_{\rm bare}
   = Z_1\,\bar h_v\Gamma\,G^{\mu\nu} h_v
   + Z_2\,v^{[\mu} v_\alpha\,\bar h_v\Gamma\,G^{\alpha\nu]} h_v \,,
\end{equation}
we observe that relation (\ref{nice}) provides for two complementary
ways to calculate the anomalous dimension of the chromo-magnetic
operator. This relation may be useful in a future calculation of
$\gamma_{\rm mag}$ beyond the one-loop order.

\section{Hadronic matrix elements}

In the HQET, hadronic matrix elements of the operators $O_n^{\mu\nu}$
in (\ref{basis}) can be parametrized in terms of invariant functions
$\phi_i(w,\mu)$, which are generalizations of the Isgur--Wise form
factor. These functions are most conveniently introduced using a
covariant formalism, in which heavy hadrons are classified in
multiplets of the heavy-quark spin symmetry and described by spin
wave functions with the appropriate transformation properties
\cite{Falk,Adam}. In particular, the ground-state pseudoscalar and
vector mesons, $P(v)$ and $V(v)$, are described by a matrix
\begin{equation}
   {\cal M}(v) = \sqrt{m_M}\,\frac{1+\rlap/v}{2}\,\Big[
   \gamma_5\,P(v) + \rlap/e\,V(v) \Big] \,,
\end{equation}
which satisfies $\rlap/v{\cal M}(v)={\cal M}(v)=-{\cal M}(v)\,
\rlap/v$. Here $m_M$ is the hadron mass, $v$ the hadron velocity, and
$e$ the polarization vector of the vector meson. The matrix elements
of the operators $O_n^{\mu\nu}$ between meson states can be obtained
from the relation \cite{FaNe}
\begin{equation}
   \langle M(v')|\,\bar h_{v'}\Gamma\,i G^{\mu\nu} h_v\,|M(v)\rangle
   = - \hbox{Tr}\Big\{ \phi^{\mu\nu}(v,v',\mu)\,
   \overline{\cal M}(v')\,\Gamma{\cal M}(v) \Big\} \,,
\label{meson}
\end{equation}
where $\mu$ is the scale at which the operators are renormalized. The
tensor form factor $\phi^{\mu\nu}(v,v',\mu)$ can be decomposed in
terms of three scalar functions, i.e.
\begin{eqnarray}
   \phi^{\mu\nu}(v,v',\mu) &=& (v^\mu v'^\nu - v^\nu v'^\mu)\,
    \phi_1(w,\mu) \nonumber\\
   &&\mbox{}+ \Big[ (v-v')^\mu\gamma^\nu
    - (v-v')^\nu\gamma^\mu \Big]\,\phi_2(w,\mu)
    + i\sigma^{\mu\nu}\,\phi_3(w,\mu) \,.
\end{eqnarray}
Evaluating (\ref{meson}) for the eigenoperators ${\cal O}_n^{\mu\nu}$
in (\ref{eigen}), we obtain the combinations of scalar functions that
are renormalized multiplicatively. They are
\begin{eqnarray}
   \psi_1(w,\mu) &=& \phi_3(w,\mu) \,, \nonumber\\
   \psi_3(w,\mu) &=& \phi_3(w,\mu) - (w-1)\,\phi_2(w,\mu) \,,
    \nonumber\\
   \psi_4(w,\mu) &=& (w+1)\,\phi_1(w,\mu) - 2\phi_2(w,\mu)
    - \phi_3(w,\mu) \,,
\end{eqnarray}
and the corresponding anomalous dimensions are $\gamma_1(w)$,
$\gamma_3(w)$, and $\gamma_4(w)$, respectively. The operator ${\cal
O}_2^{\mu\nu}$ has a vanishing matrix element between the
ground-state mesons.

In leading logarithmic approximation, the scale dependence of the
functions $\psi_n(w,\mu)$ is the same as that of the operators ${\cal
O}_n^{\mu\nu}$ shown in (\ref{LO}). The theoretical predictions for,
e.g., weak decay form factors in the HQET involve these functions
renormalized at the large scale set by the mass of the heavy quark
that decays. Our results can then be used to rewrite the results in
terms of functions renormalized at a low scale $\mu\ll m_Q$, which
may be identified with the scale at which a non-perturbative
evaluation of these functions is performed. As an example, consider
the combination $f(w,m_b)=\psi_1(w,m_b)+2\psi_3(w,m_b)$, which
parametrizes a class of non-factorizable contributions in
non-leptonic weak decays such as $\bar B^0\to D^+\pi^-$ \cite{Blok}.
Using our results, we can relate the function $f(w,m_b)$ with the
functions $\psi_n(w,\mu)$ renormalized at a low scale $\mu\approx
1$~GeV, which have been calculated recently using QCD sum rules
\cite{MN96}. The result is
\begin{equation}
   f(w,m_b) = \left( \frac{\alpha_s(m_b)}{\alpha_s(\mu)}
   \right)^{\gamma_1^0(w)/2\beta_0} \psi_1(w,\mu)
   + 2 \left( \frac{\alpha_s(m_b)}{\alpha_s(\mu)}
   \right)^{\gamma_3^0(w)/2\beta_0} \psi_3(w,\mu) \,.
\end{equation}

The functions $\psi_n(w,\mu)$ obey non-trivial normalization
conditions at $w=1$. They are given by \cite{FaNe}
\begin{equation}
   \psi_1(1,\mu) = \psi_3(1,\mu) = \lambda_2(\mu) \,, \qquad
   \psi_4(1,\mu) = - \frac 23\,\lambda_1 \,,
\label{psirel}
\end{equation}
where $\lambda_1$ and $\lambda_2(\mu)$ parametrize the forward matrix
elements of the kinetic and chromo-magnetic operators present in the
effective Lagrangian (\ref{Leff}). The anomalous dimension
determining the scale dependence of the parameter $\lambda_2(\mu)$ is
$\gamma_{\rm mag}$, whereas the parameter $\lambda_1$ has no
(logarithmic) scale dependence because of reparametrization
invariance \cite{LuMa}. Thus, the relations (\ref{psirel}) are in
accordance with our exact results in (\ref{gamZ}) and (\ref{gamel}).

\section{Conclusions}

We have discussed the general structure of the mixing and
renormalization of velocity-changing operators of the type $\bar
h_{v'}\Gamma\,G^{\mu\nu} h_v$ in the HQET. The primary motivation for
this study is the connection between these operators and the
velocity-conserving operators appearing at order $1/m_Q$ in the
effective Lagrangian of the HQET. Besides, the hadronic matrix
elements of such operators determine part of the $1/m_Q^2$
corrections to heavy-hadron weak decay form factors, as well as a
class of non-factorizable corrections in non-leptonic two-body decays
of $B$ mesons.

We have constructed the combinations of operators that are
renormalized multiplicatively, and calculated the corresponding
velocity-dependent anomalous dimensions at the one-loop order. We
have also considered the hadronic matrix elements of the operators
$\bar h_{v'}\Gamma\,G^{\mu\nu} h_v$ and determined, for the case of
the ground-state heavy mesons, the scale dependence of the
corresponding hadronic form factors. In the limit of equal
velocities, we find that the chromo-electric and the chromo-magnetic
operators do not mix under renormalization. Moreover, as a
consequence of the virial theorem and reparametrization invariance,
the anomalous dimension of the chromo-electric operator vanishes to
all orders in perturbation theory. This implies an exact relation
between renormalization constants, which may help in the calculation
of the yet unknown two-loop anomalous dimension of the
chromo-magnetic operator.

\vspace{0.3cm}
{\it Acknowledgements:\/}
G.A.\ acknowledges a grant from the Generalitat Valenciana. He also
likes to acknowledge the hospitality of the CERN Theory Division,
where this research was carried out.

\end{document}